\documentclass{eas}
\usepackage{graphicx}

\begin{document}

\title{The Stability and Prospects of the 
Detection of Terrestrial/Habitable Planets in Multiplanet 
and Multiple Star Systems } 

\runningtitle {Stability and Prospects of the Detection of 
Terrestrial Planets}

\author{Nader Haghighipour}
\address{Institute for Astronomy and NASA Astrobiology Institute,
University of Hawaii-Manoa, Honolulu, HI 96822, USA, 
nader@ifa.hawaii.edu}

\begin{abstract}
Given the tendency of planets to form 
in multiples, and the observational evidence
in support of the existence of 
potential planet-hosting stars in binaries or clusters,
it is expected that extrasolar terrestrial planes 
are more likely to be found in
multiple body systems. This paper discusses the 
prospects of the detection of terrestrial/habitable planets in 
multibody systems by presenting the results of a study 
of the long-term stability of these objects in systems with multiple 
giant planets (particularly those in eccentric and/or in mean-motion 
resonant orbits), systems with close-in Jupiter-like bodies, and
systems of binary stars. The results of simulations 
show that while short-period terrestrial-class objects 
that are captured in near mean-motion resonances with migrating giant 
planets are potentially detectable via transit photometry
or the measurement of the variations of the transit-timing due 
to their close-in Jovian-mass planetary companions, the prospect of 
the detection of habitable planets with radial velocity 
technique is higher in systems with multiple giant planets 
outside the habitable 
zone and binary systems with moderately separated stellar companions. 
\end{abstract}

\maketitle

\section{Introduction}
A survey of the currently known extrasolar planetary systems
indicates that more than 30 of these systems are hosts to two
or more giant planets. Similar to our solar system, where the
simulations of planet formation show that terrestrial and giant
planets form in multiples, many of these systems may also harbor 
smaller terrestrial-class objects. A study of the stars of these
systems reveals that some of these stars are members of binaries
or multi-star systems. In fact, approximately 25\% of 
planet-hosting stars are within dual-star environments implying that
planet formation in multi-star systems is robust and many of
these systems may host additional smaller objects.

At present, the detection of terrestrial/habitable planets in systems
with multiple giant planets and/or close stellar companions
is not a feasible task. However, space-based telescopes such as
NASA's Kepler and SIM are on their ways to pave the road to the 
discovery of many of these 
objects in the near future. It, therefore, proves useful to study the
characteristics of multibody systems, and to identify the dynamical 
and orbital properties of those that are capable of hosting
habitable planets. 

In a multibody system, the perturbation of the giant
planets and stellar companions have profound effects on the
formation and long-term stability of habitable planets.
As shown by the simulations of the formation of terrestrial planets
in our solar system, the formation of these objects (especially in the
habitable zone) and their water contents are strongly affected 
by the dynamics of Jupiter (Chambers \& Cassen \cite{Chambers02}). 
The perturbative effect of this object
will increase the orbital eccentricities of the planetesimals
and protoplanetary bodies, which in turn results in their collision and/or
ejection from the system. The radial mixing of these objects,
induced by the close approach of Jupiter, delivers planet-forming material,
including water-carrying protoplanets,
from different regions of the asteroid belt to the accretion zone of
proto-Earth.

The efficiency of the removal of material and the collisional growth
of protoplanetary bodies is a function of the orbital eccentricity
of the giant planet. For instance,
as shown by Chambers \& Cassen (\cite{Chambers02}) 
and Raymond (\cite{Raymond06}), an eccentric
Jupiter will cause severe depletion of icy asteroid from the outer
part of the asteroid belt, which subsequently results in the formation
of a smaller and drier Earth.
In an extrasolar planetary system with multiple
stars and/or giant planets, 
such effects are much stronger. In order for such multibody systems to 
be habitable, not only do Earth-like planets, with substantial amounts of
water, have to form in their habitable zones,
the orbits of these objects have to be immune to the perturbation
of other bodies and maintain their stability for long times.

This paper presents the results of an extensive study in identifying
systems in which the interactions among multiple bodies
allow long-term stability of terrestrial-class objects and 
the indirect detection of these bodies. In the next section, the 
results of a study of the stability of habitable planets in
system with multiple giant planets are presented. As examples
of such systems, the two systems of 55 Cnc and GJ 876 are discusses.
In section 3, habitability of moderately close binary star
systems are studied, and in section 4 the results of
a study of the implications of the dynamics of terrestrial-class 
and Super-Earth objects on their detections using the transit timing
variation method in the vicinity of hot-Jupiters are presented.

\begin{figure}
\centering
\includegraphics[height=4.5cm]{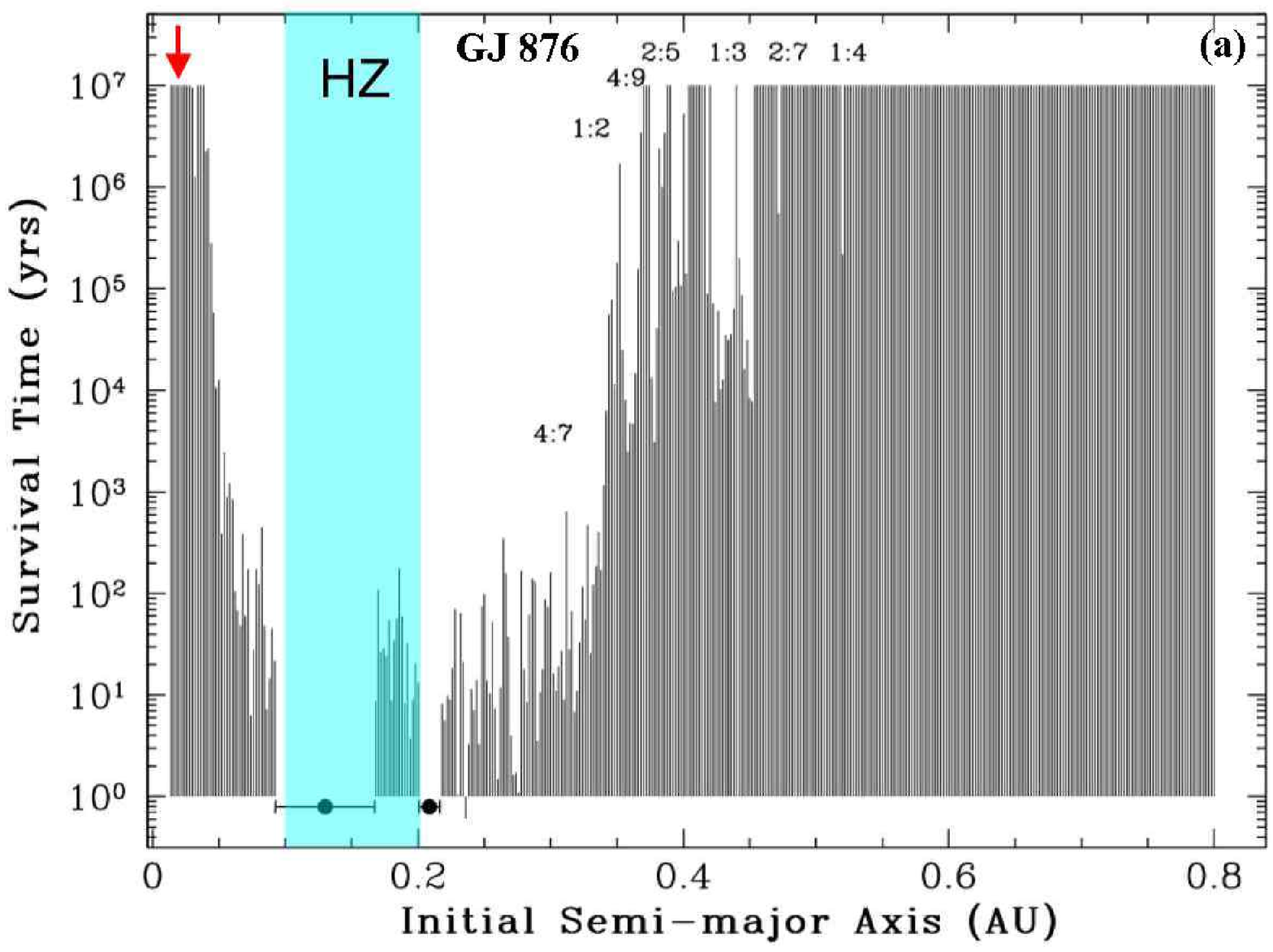}
\includegraphics[height=4.5cm]{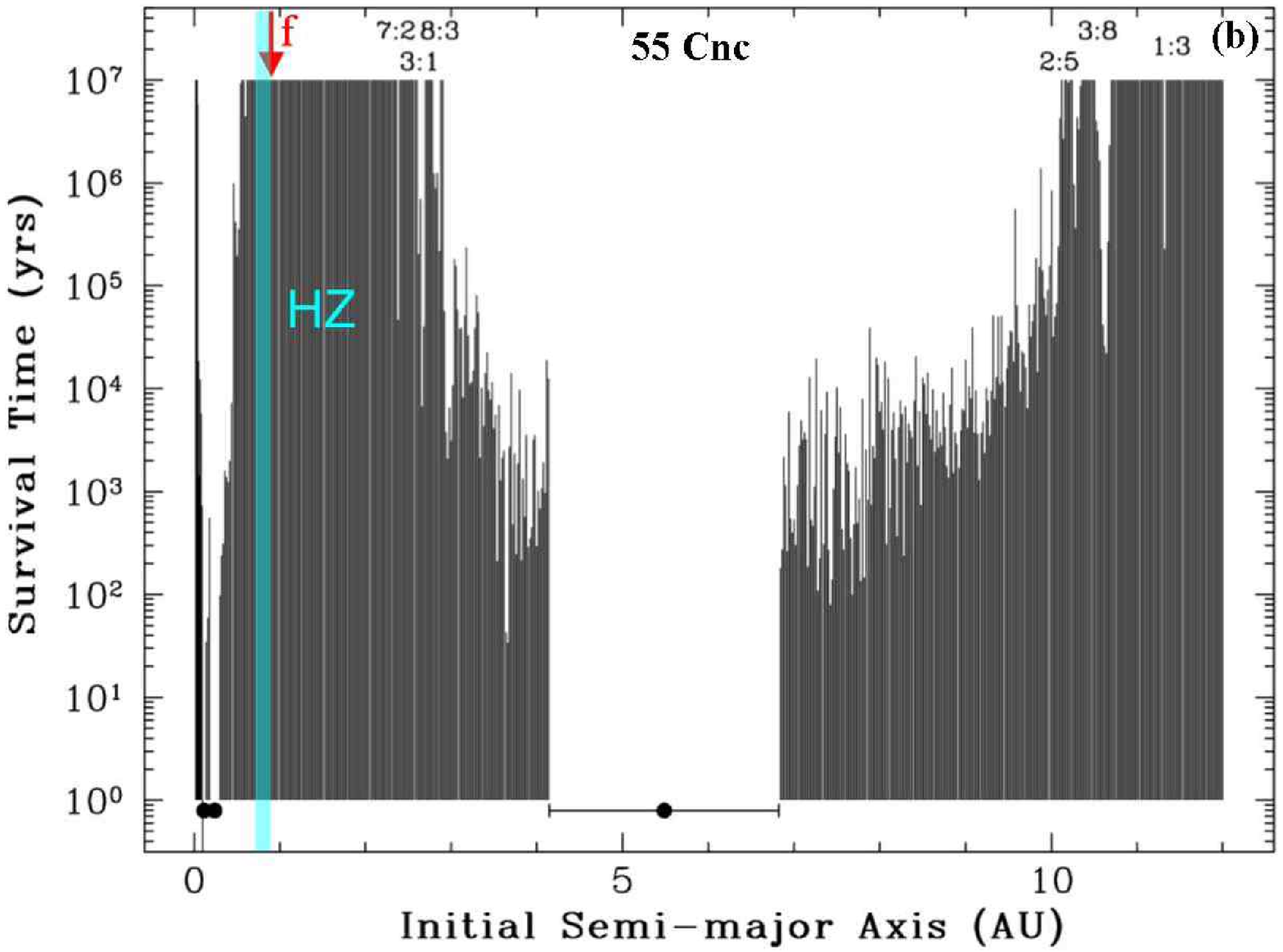}
\caption{Graphs of the lifetimes of test particles in the planetary
systems of GJ 876 and 55 Cnc (Rivera \& Haghighipour \cite{Rivera07}). 
The habitable zones of the two
systems are also shown. The red arrows in both graphs show the
locations of the terrestrial-size planets of the two systems.}
 \label{fig1}
\end{figure}

\section{Multiple Giant Planet Systems}

The formation and stability of terrestrial planets are strongly
affected by the perturbation of giant bodies.
Not only do the latter objects remove protoplanetary material that 
may contribute to the formation of terrestrial planets, they
may also destabilize the regions were these planets
can form. Such perturbative effects are stronger in extrasolar planetary 
systems with several Jovian-type bodies, in particular those in
eccentric orbits (e.g., GJ 876, $\upsilon$ Andromedae, 47 UMa, 
55 Cnc, Gl 581, HD 40307, HD 37124, HD 74156, HD 160691, and HD 69830).

Prior to the study of habitable planet formation in such systems, it
is necessary to determine whether the orbit of an Earth-like planet
in the habitable zones of their central stars are stable. 
For this purpose, the equations of the motion of an Earth-like planet 
have to be integrated for a long time and for different values of its 
orbital elements. Given the system's vast 
parameter-space, performing such integrations will be a very
time-consuming task. However, it is possible to consolidate 
many of these integrations into one by taking into account that
an Earth-like planet is two orders of magnitude smaller than a giant
body and its affect on the orbital dynamics of the latter 
object may be negligible. In other words, to the zeroth order of 
approximation, Earth-like 
planets can be replaced by test particles. Given that it is desired
for a habitable planet to spend most of its orbital motion in
the system's habitable zone, it would also not be too unrealistic to 
assume that these test particles have circular orbits. Such test
particle approximations are not exact, but produce a 
rough mapping of the parameter-space which indicates regions
where small planets can 
have stable orbits. Figure 1 shows the results of two of such 
integrations for GJ 876 and 55 Cnc (Rivera \& Haghighipour \cite{Rivera07}). 
GJ 876 is a system of two giant planets with semimajor axes of 0.13 AU 
and 0.21 AU, and eccentricities of 0.27 and 0.025, respectively. The
motion of an Earth-like planet in the habitable zone
of this system (0.1-0.25 AU) is disturbed by the gravitational
forces of its two giant bodies. 
Results of the numerical integrations of test particles 
indicate that the habitable zone of this system is unstable
and as expected, stability occurs at distances 
outside the influence zones of the giant planets. The stable regions of 
Figure 1a are representatives of orbits in which Earth-size planets may 
potentially exist. As an example,  
the location of the recently discovered 7.5 Earth-mass planet
of this system (Rivera {\em et al.\/} \cite{Rivera05}), 
denoted by a red arrow in Figure 1a, is in one of such stable regions.

The planetary system of 55 Cnc shows a different characteristic. 
Unlike GJ 876, this system has a habitable zone that extends from 0.70 AU
to 2.71 AU and is outside the gravitational influence of the system's 
three giant planets ($a$=0.115, 0.24, 5.77 AU, and 
$e$= 0.014, 0.086, 0.025). 
In this system, the habitable zone and a vast area in its vicinity are
stable and can host smaller (i.e., terrestrial-class) objects. 
The two Earth-size planet of this system, with semimajor axes of 
0.038 AU and 0.781 AU, are located in this area
(Fischer {\em et al.\/} \cite{Fischer08}). These integrations indicate 
that, in a multiplanet system, although small planets may exist at
distances outside the orbits of giant bodies, 
terrestrial-size objects are more
likely to be found in systems where giant planets are so far away from
the central star that stable regions exist interior to their orbits.
For more details, the reader is referred to
Rivera \& Haghighipour (\cite{Rivera07}).

\section{Binary Star Systems}

Other interesting and dynamically complicated multibody 
planetary systems are planets in binary stars. In such systems,
(giant) planets revolve around a star of a binary. 
Approximately 25\% of the currently known stars with
extrasolar planets are of this kind. The fact that (giant) planets 
exist in such environments is an indication that planet formation 
in dual-star systems is robust and many of these systems may also 
host smaller (e.g., terrestrial-class) objects.  

The formation and the prospect of the detection of habitable planets 
in {\it binary-planetary} systems are affected by the perturbations of their
giant planets and the gravitational force 
of their stellar companions. In the majority of the currently known
binary-planetary systems, the stellar companion is at distances of
250 AU to 6000 AU from the primary (planet-hosting) star
and its perturbative effect is negligible.
However, systems with giant planets exist (namely,  GJ 86 and $\gamma$ Cephei)
in which the binary separation is smaller than 20 AU. In such systems,
terrestrial/habitable planet formation is strongly affected by both 
the giant planet and the secondary star. Simulations of the orbital
evolution of an Earth-like planet in the $\gamma$ Cephei system
have shown that the rate of the survival of a terrestrial planet
is much higher at distances close to the parent star. In other words,
habitable planets can have stable orbits in such 
close binaries when the habitable zone is outside the region
of the influence of the giant
planet and close to the planet-hosting star 
(Haghighipour \cite{Haghighipour06}).

The above-mentioned stability condition categorizes the type
of the close binary-planetary systems that may be habitable, and raises 
the question of the possibility of the formation of Earth-like objects 
in the habitable zones of such systems. A recent study by 
Haghighipour \& Raymond (\cite{Haghighipour07}) 
has addressed this issue by simulating the late stage of terrestrial 
planet formation around the primary of a moderately close binary-planetary
system. The left graph of Figure 2 shows 
snap shots of the formation and evolution of one of such simulations. 
The system in this figure consists of two Sun-like stars
with a separation of 30 AU. The binary eccentricity is 
0.2 and the primary is host to a Jupiter-size planet in a circular 
orbit at 5.2 AU. The system also includes a disk of protoplanetary 
bodies with 120 Moon- to Mars-sized objects distributed randomly 
between 0.5 AU and 4 AU. The orbit of the giant planet was chosen 
to be circular to minimize its initial perturbative effect on the 
collision and evolution of protoplanetary objects. To simulate the 
delivery of water to the region of the accretion of a terrestrial 
planet around the primary star, a water gradient was assumed for 
the protoplanetary bodies. Similar to the water distribution among  
the primitive asteroids in the asteroid belt, inside 2 AU, 
protoplanets were taken to be dry, between 2 AU and 2.5 AU, 
they were given 1\% water, and beyond 2.5 AU, their water content 
was set to 5\%. Simulations show that after 100 Myr, an Earth-size object
(1.17 Earth-mass) was formed inside the habitable zone at 1.16 AU,
with an eccentricity of 0.02 and water content of 0.164\%. 
As a point of comparison, the orbital eccentricity of Earth is 
0.017 and its water content is approximately 0.1\%.

\begin{figure}
\centering
\includegraphics[height=4.5cm]{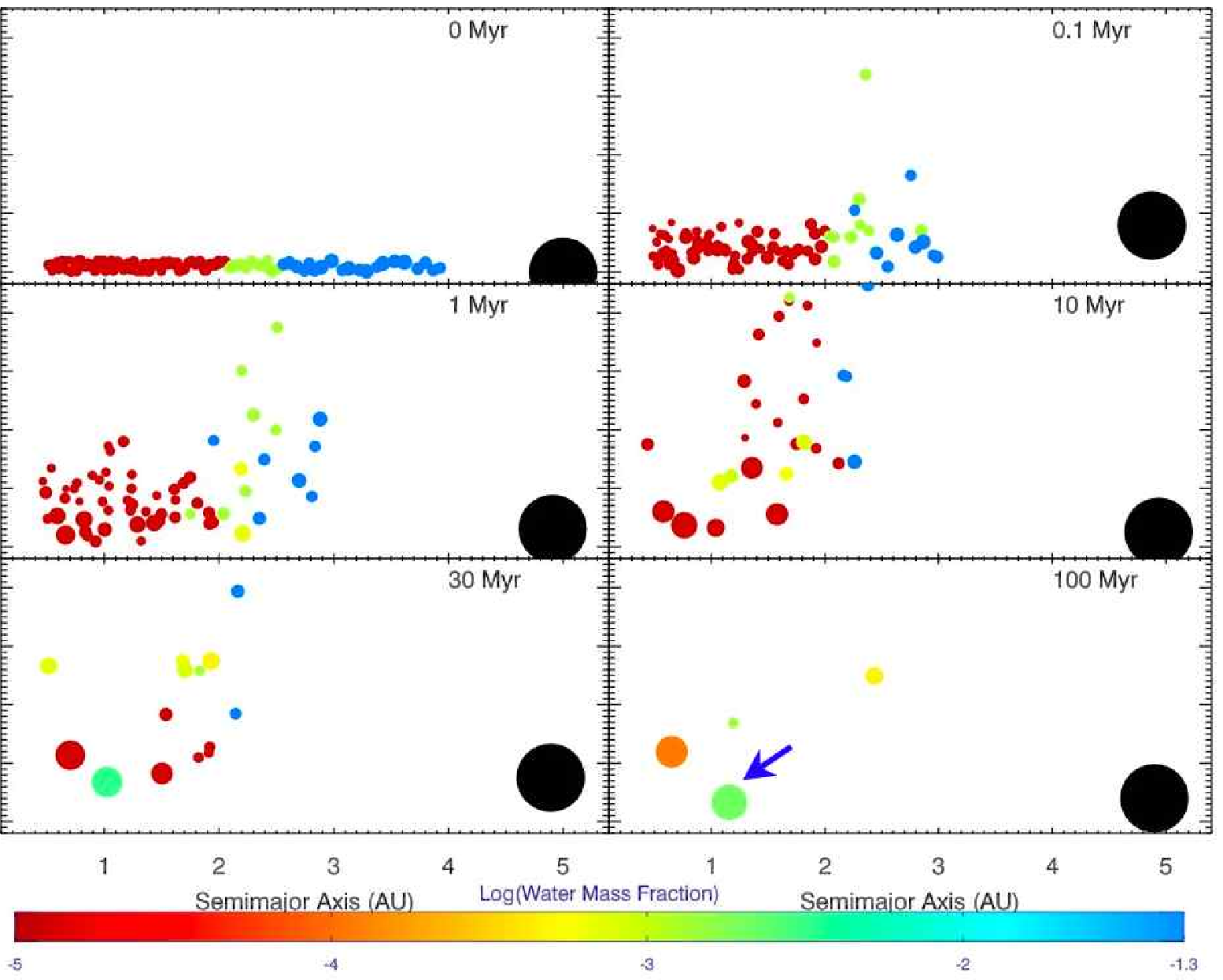}
\includegraphics[height=4.5cm]{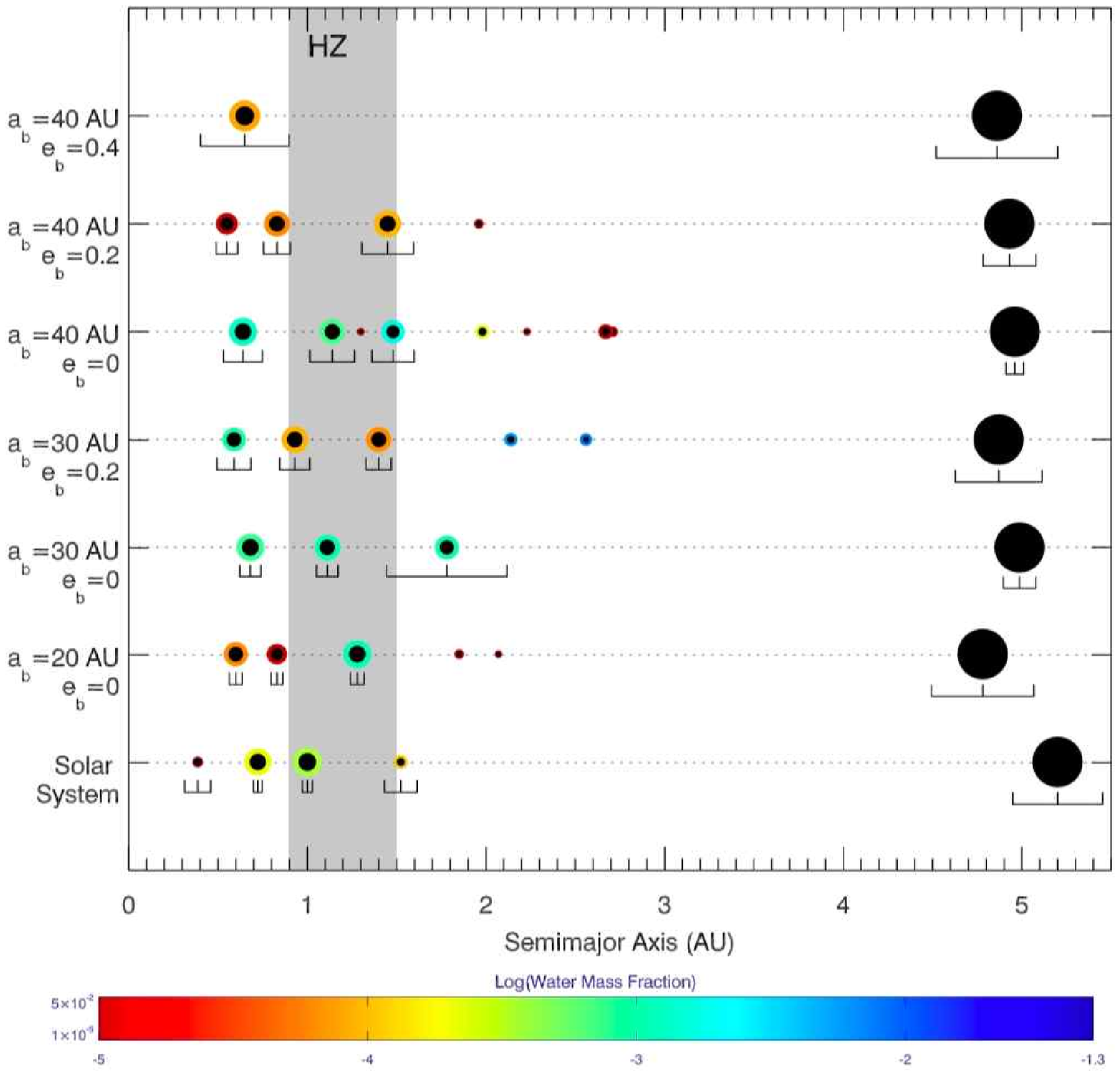}
\caption{Left: Snap shots of the late stage of 
terrestrial planet formation in a 30 AU binary. The black circle 
represents the giant planet of the system. The primary star is at 
the origin and the secondary star is not shown. Right: The final
assembly of planets for different values of the mass-ratio, semimajor
axis and eccentricity of the binary. Graphs from Haghighipour \& Raymond
(\cite{Haghighipour07}).}
 \label{fig2}
\end{figure}

Simulations were also carried out for different values of the
binary's mass-ratio, semimajor axis, and orbital eccentricity.
The right graph of
Figure 2 shows some of the final results for a binary mass-ratio of
unity. As shown here, the sizes and water contents of the final
terrestrial planets vary with the semimajor axis and eccentricity 
of the stellar companion. In fact, it is the periastron distance
of the secondary star that dictates the evolution of the system and the
final assembly of the terrestrial planets. In binaries with small
periastra, the interaction between the secondary star and the 
giant planet increases the orbital eccentricity of the latter body
and causes close approaches between this object and the planetary embryos.
By transferring angular momentum from the secondary star to the
protoplanetary disk, the giant planet causes many of these object,
including those that carry water, to be ejected from the system. 
As a result, in closer and eccentric binaries, the final products
of the simulations are smaller terrestrial planets that contain 
less or no water. The left graph of Figure 3 shows the relation 
between the periastron of the binary $(q_{\rm b})$ and the
semimajor axis of the outermost terrestrial planet $(a_{\rm out})$.
As shown in this graph, similar to 
Quintana {\em et al. \/} (\cite{Quintana07}),
simulations with no giant planets favor regions interior to 
$0.19{q_{\rm b}}$ for the formation of terrestrial objects.
This implies that, for the simulation of Figure 2, where the 
primary is a Sun-like star, and the inner edge of the habitable
zone is at $\sim 0.9$ AU, habitable planets would not form
if the perihelion distance of the secondary star is smaller
than 0.9/0.19 = 4.7. In simulations with
giant planets, on the other hand, Figure 3 shows that 
terrestrial planets form closer-in. The ratio 
${a_{\rm out}}/{q_{\rm b}}$ 
in these systems is between 0.06 and 0.13.

A detailed analysis of the simulations of the systems in
which habitable planets formed indicates that binaries in 
these systems have large periastra. The right graph of Figure 3
shows this for the simulations of Figure 2. The circles in this 
figure represent systems with habitable planets.
The numbers on the top of the circles show the mean 
eccentricity of the giant planet at the end of each simulation.
As a comparison, systems 
with unstable giant planets have also been marked. Since at the 
beginning of each simulation, the orbit of the giant planet 
was considered to be circular, a non-zero eccentricity for the 
orbit of this object is the result
of the interaction between this body and the secondary star.
As shown in this graph, Earth-like objects are formed in systems 
where the orbital eccentricity of the giant planet is small.
In other words, systems in which the interaction between this
object and the secondary star is weak. Figure 3 indicates that
habitable planet formation favors binary-planetary  systems 
with  moderate to large
perihelia, and with giant planets on low eccentricity orbits.

\begin{figure}
\vskip 20pt
\centering
\includegraphics[height=3.8cm]{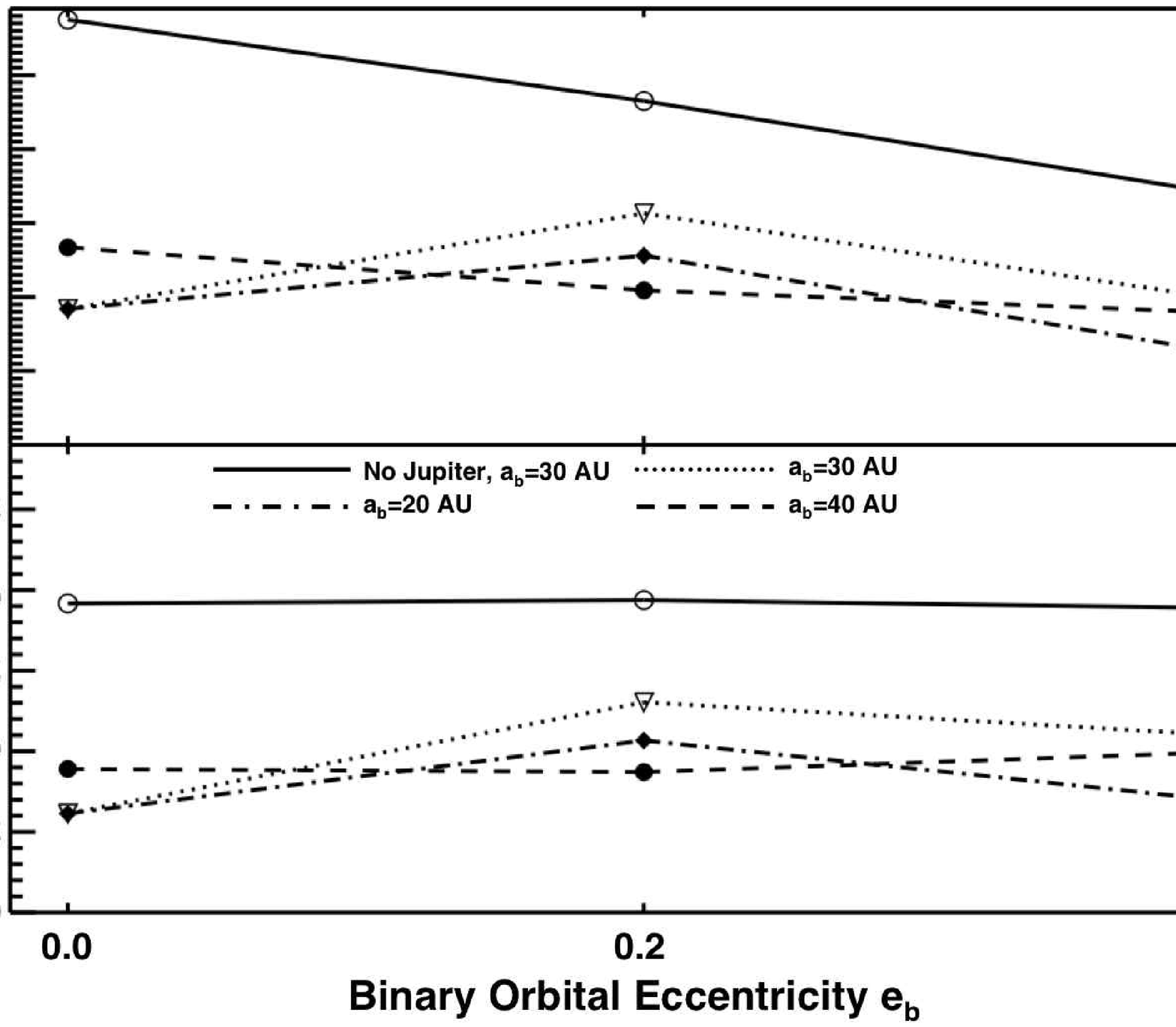}
\includegraphics[height=3.8cm]{Haghighipour-f3b.eps}
\caption{Left: Graph of the relation between the perihelion
of an equal-mass binary and the location of its 
outermost terrestrial planet. 
Right: The region of the $({e_b},{a_b})$ space
for a habitable binary-planetary system. Graphs from Haghighipour \& Raymond
(\cite{Haghighipour07}).}
 \label{fig3}
\end{figure}

\section{Transiting Systems}

Detecting more than 50 extrasolar planets during the past decade, 
transit photometry has proven to be a successful technique in
identifying planetary bodies around other stars. This technique
that measures the decrease in the brightness 
of a star when occulted by a transiting planet, is more sensitive 
to close-in and larger objects; the currently known transiting planets
are of Neptune- to Jupiter-size and majority of them have orbital 
periods between 2 to 4 days.

It is widely accepted that such close-in giant planets were formed at 
large distances and arrived at their current 
orbits after undergoing radial migration through interaction with the 
circumstellar disks of their parent stars. 
Numerical simulations have indicated that, despite the scattering of
planet-forming material and fully formed planets to large distances
during the inward migration of a giant body,
Earth-like planets can still form in such systems
and possibly close to the transiting object 
(Raymond {\em et al.\/} \cite{Raymond06b}, Mandell {\em et al.\/} 
\cite{Mandell07}). In the case of the latter,
the gravitational force of the smaller planet perturbs the orbit of the
giant body and causes variations in the time of its transits.
As shown by Agol {\em et al.\/} (\cite{Agol05}) and 
Steffen \& Agol (\cite{Steffen05}), such transit timing 
variations (TTV) can be used to identify the orbital and physical 
characteristics of the perturbing terrestrial planet.

The variations in the transit timing of a giant planet 
are functions of the mass and orbital elements 
(i.e., semimajor axis and eccentricity) of the
perturbing body. Agol {\em et al.\/} (\cite{Agol05}) 
have shown that these variations become considerably
large when the perturbing and transiting planets are in a mean-motion
resonance. Given the small size of a perturbing object 
(i.e., terrestrial-size) and
its proximity to the transiting body, it would not be surprising if
for some values of the orbital elements of the two planets, 
the orbit of the former object became unstable. Since large TTV 
signals require strong interactions between the transiting and perturbing 
planets (i.e., close approaches which are results of larger orbital 
eccentricity), and they are larger when the two bodies
are in a mean-motion resonance, it is therefore necessary 
to identify regions of the parameter-space for which an Earth-like planet
can maintain its orbit for long times and produce large TTV signals.

\begin{figure}
\centering
\includegraphics[height=8cm]{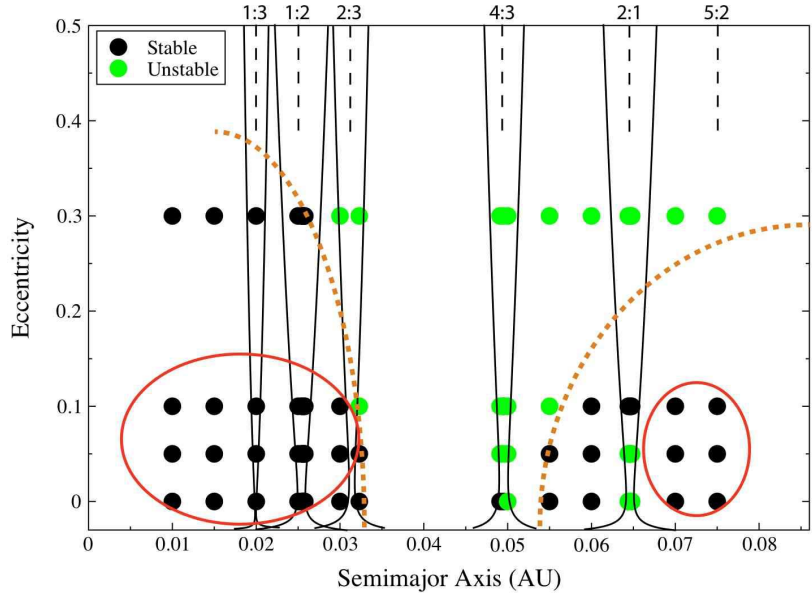}
\caption{Graph of stability for an Earth-size planets in a
transiting system. The central star is Sun-like.
The transiting planet (not shown here) is Jupiter-like in a 3-day circular
orbit at 0.0407 AU. The circles
correspond to the values of the semimajor axis and eccentricity
of the Earth-like planet at the beginning of each simulation. The orange 
dotted lines show the boundaries of the unstable region for an Earth-size 
planet. The red
circles present conservative regions of the stability of a 
terrestrial-class object.}
 \label{fig4}
\end{figure}

Recently Haghighipour {\em et al.\/} (\cite{Haghighipour08}) 
have addressed this issue by integrating the motion of
an Earth-size planet in a transiting system. The planetary model considered
by these authors consists of a Sun-like star, a transiting Jupiter-size 
planet in a 3-day circular orbit, and an Earth-size object. Using the N-body 
integration package MERCURY (Chambers \cite{Chambers99}), 
these authors simulated the dynamics 
of this system for different values of the semimajor axis (0.01 AU to
0.075 AU), orbital eccentricity (0, 0.05, 0.1, 0.3), and orbital inclination
(0, 5 deg.) of the small body for 1 Myr. Results for a co-planar system 
are shown in Figure 4. The circles in this figure represent the values of
the semimajor axis and eccentricity of the Earth-size planet at the
beginning of each simulation. Black circles correspond to stable orbit
whereas green indicates instability. The widths of the major resonances are
also shown. Note that no Earth-like planet was placed in the region between
$a-3{R_H}$ and $a+3{R_H}$ where $a$ and $R_H$ are the semimajor axis and
Hill radius of the Jupiter-like planet. As shown in this figure, regions
exist where the orbit of the Earth-size planet becomes unstable. For outer
orbits, this region extends to the outer edge of 2:1 MMR and is limited
to eccentricities below 0.2. For the inner orbits, stability seems to
exist interior to 2:3 MMR and 
for the values of eccentricity equal to and smaller than 0.3. 
Integrations for longer times (beyond 1 Myr) and for a massive perturber
(up to 15 Earth-mass) have, however, indicated that, except for 1:2
and 1:3 MMRs, where the orbit of the Earth-like planet stays stable for
values of its orbital eccentricity up to 0.5 
(Haghighipour \cite{Haghighipour08b}), as the mass of
the Earth-like planet increases, the boundaries of the unstable region
expand and tend toward lower values of eccentricity. In a conservative
approach, the red circles in Figure 4 show the most stable
regions of the eccentricity-semimajor axis space. As shown here,
a terrestrial-class object in the vicinity of a transiting giant planet,
will be stable if its orbit has an eccentricity lower than 0.2 and
is in a low-order MMR with the giant body. As shown in Figure 5, the
TTV signal produced by such a perturber is quite large and falls within
the sensitivity of the current detection techniques.

\begin{figure}
\centering
\includegraphics[height=4cm]{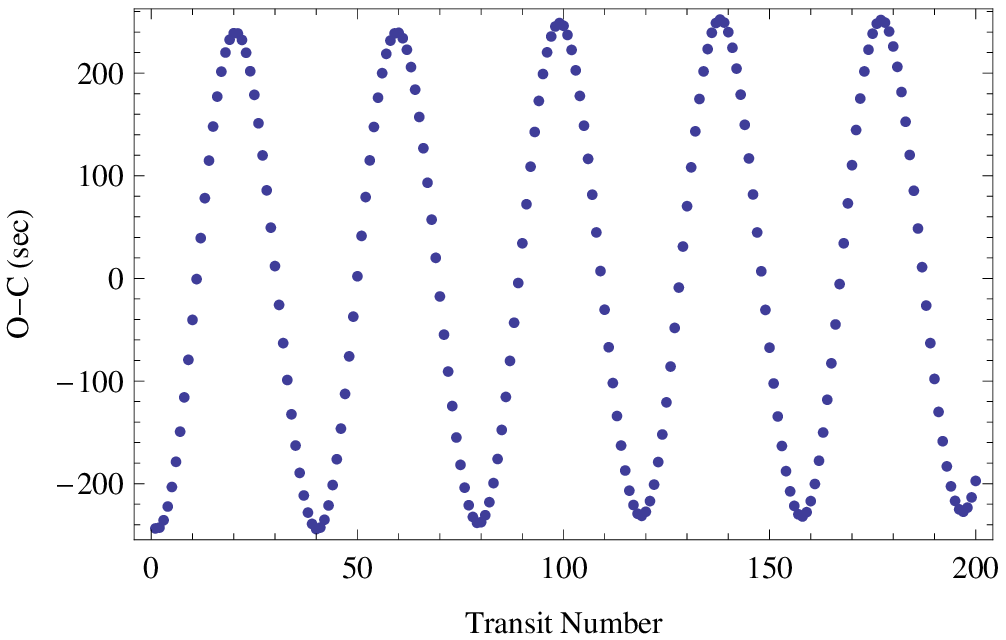}
\includegraphics[height=4cm]{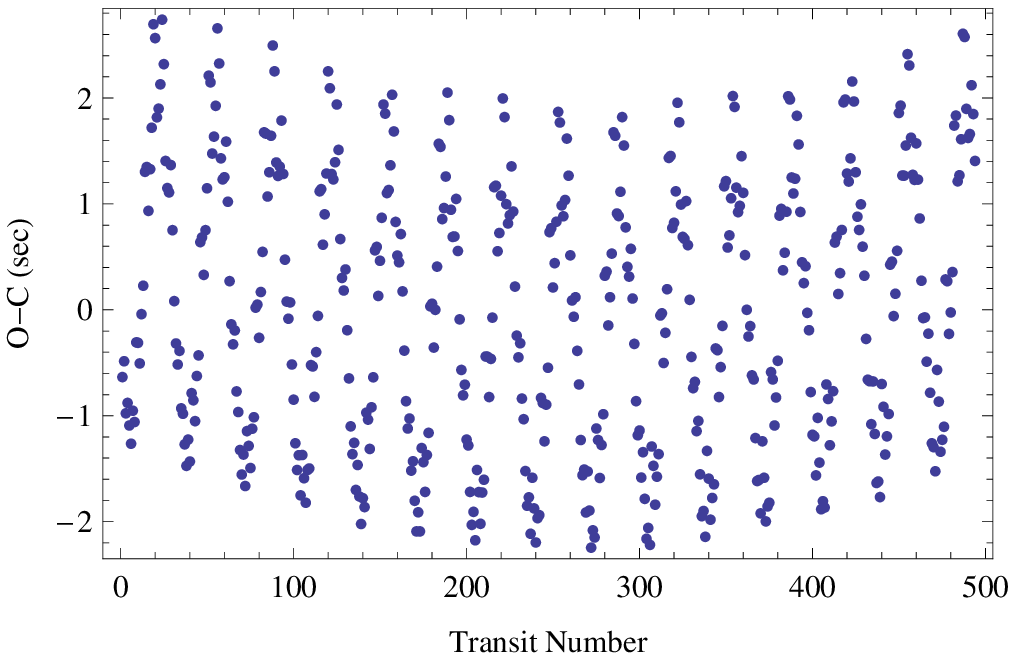}
\caption{Graphs of TTV signals for an Earth-size planet in an
interior 1:2 (Left) and exterior 5:2 (right) MMR with a transiting 
giant planet. Graphs from Haghighipour {\em et al.\/} 
(\cite{Haghighipour08c})}.
 \label{fig5}
\end{figure}

\vskip 10pt
This work was partially supported by 
the office of the Chancellor of the University of Hawaii, and a 
Theodore Dunham Jr. grant administered by Funds for Astrophysics 
Research, Inc., and
the NASA Astrobiology Institute 
under Cooperative Agreement NNA04CC08A at the Institute for Astronomy
at the University of Hawaii.
Access to the computational facilities of the
Department of Terrestrial Magnetism at the Carnegie Institution of 
Washington where some of the simulations of this paper were performed,
is gratefully acknowledged.


\begin{thebibliography}{99}
\bibitem[2005]{Agol05} Agol, E., Steffen, J., Sari, R. \& Clarkson, W.
2005, MNRAS, 359, 567.
\bibitem[1999]{Chambers99} Chambers, J.E. 1999, MNRAS, 304, 793.
\bibitem[2002]{Chambers02} Chambers, J.E. \& Cassen, P. 2002, 
M\&PS, 37, 1523.
\bibitem[2008]{Fischer08} Fischer, D.A. \etal\ 2008, ApJ, 675, 790.
\bibitem[2006]{Haghighipour06} Haghighipour, N. 2006, ApJ, 644, 543.
\bibitem[2007]{Haghighipour07} Haghighipour, N. \& Raymond, S.N. 2007,
ApJ, 666, 436.
\bibitem[2008]{Haghighipour08} Haghighipour, N., Agol, E. \& Steffen, J.
2008, BAAS, 39, 768.
\bibitem[2008c]{Haghighipour08c} Haghighipour, N., Steffen, J. \&
Agol, E. 2008, Astrobiology, 8, 374. 
\bibitem[2008b]{Haghighipour08b} Haghighipour, N., Steffen, J. \&
Agol, E. 2008, Am.Geophys.Soc., P14B-04. 
\bibitem[2007]{Mandell07} Mandell, A.M., Raymond, S.N. \& Sigurdsson, S.
2007, ApJ, 660, 823.
\bibitem[2007]{Quintana07} Quintana, E.V., Adams, F.C., Lissauer, J.J.
\& Chambers, J. E. 2007, ApJ, 660,807.
\bibitem[2006]{Raymond06} Raymond, S.N. 2006, ApJ, 643, L131.
\bibitem[2006]{Raymond06b} Raymond, S.N., Mandell, A.M. \& Sigurdsson, S.
2006, Science, 313, 1413. 
\bibitem[2005]{Rivera05} Rivera, E. \etal\ 2005, ApJ, 634, 625.
\bibitem[2007]{Rivera07} Rivera, E. \& Haghighipour, N. 2007, MNRAS, 374, 599.
\bibitem[2005]{Steffen05} Steffen, J. \& Agol, E. 2005, MNRAS,
364, L96.


\end{thebibliography}
\end{document}